\documentclass[reqno,dvips,12pt]{article} 

\usepackage{epsfig,graphicx,amsmath,amsfonts,verbatim,amssymb,amsthm}
\theoremstyle{plain}
\newtheorem{theorem}{\qquad Theorem }

\theoremstyle{remark}
\newtheorem{remark}{\qquad Remark}

\begin{document}

\title{\textbf{On Refinement of Several Physical Notions
and Solution of the Problem of Fluids for Supercritical States}}
\author{\textbf{V.~P.~Maslov}
}
\date{ }

\maketitle

\begin{abstract}

To solve the ancient problem of fluids,
i.e., of states in which  there is no difference between gas and liquid
(the so-called supercritical states),
it is necessary to abandon several ``rules of the game'',
which are customary to physicists
and to refine them by using rigorous mathematical theorems.
\end{abstract}

\hspace{4cm} To the memory of V.~L.~Ginzburg

\section{Introduction}

Physicists and mathematicians have different language
and different logics. Namely, a mathematician first formulates
the conditions, then the theorem (the formula), and only after this proves it.
A physicist, conversely, first derives a formula, which he makes rather
convincingly, but he often conceals the conditions assuming them to be obvious.
As a rule, mathematicians prove theorems,
which have already been ``known'' to physicists
in the sense that the physicists have been using
them explicitly or implicitly for a long time.

There is a series of rather well-known theorems
bearing the names of their authors,
which were formulated in a different language
and, in fact, had been already proved by the physicists\footnote{Proved
in the sense that neither a rigid statement nor a rigid proof
required any efforts.}.
The results of mathematicians who are translators
to the mathematical language were especially easily and rapidly
understood by the mathematical community.

Sometimes, a mathematician, who also obtained new results in physics
and, in addition, tried to explain them in the physical language,
was not taken as a mathematician by the mathematical community,
and he had to prove this separately.

The author also happened to obtain several results concerning the area
of interests of those who doubted\footnote{That the  author
is a rigorous mathematician and it is impossible to find errors
in his works.},
see~\cite{UMN_23_2},~\cite{FAN_2_2}, and the work~\cite{UMN_14_4}
continued in~\cite{RJ_1_1} (how Levsha had to ``show a flea''),
in order to make the mathematical community to acknowledge him.

Similarly, it is difficult to make the physicists to believe
that the mathematicians, who follow rigorous logics, can obtain
results that are paradoxical from the point of view of customary
understanding. Physicists are not convinced even by the fact that
the mathematical results correspond to well-known experiments
and predict new experiments, which are in a rather good agreement
with the theoretical calculations.
One of my friends, a famous physicist, used to tell me:
``You do not convince me by your lemmas''.

In the second part of the famous book of Feynman and Hibbs
about continual integrals, the unrigorous logics of Feynman continuous integrals
(which was useful for physicists) led to the rigorous results obtained
by Norman Wiener twenty years ago (no reference was made).

The author had to ``prove'' the semiclassical asymptotics
of the Feynman continuous integrals
(he proved it indirectly somewhat earlier) according to the Feynman logics
(V.~Guillemin and S.~Sternberg repeated the proof in~\cite{Guillemin},
but fairly quoted the word ``proof'').

The ``rules of the game'' developed by physicists
are very useful, because they allow them to ``jump over''
the mathematical difficulties and hence to proceed significantly faster.
Their physical intuition developed in experiments in different areas of physics
helps them to feel the result empirically, especially if the numerical data
are known. The latter allows them to avoid complicated asymptotics
in several small parameters and a nonstandard analysis.
In the first aspect (the intuition), I was always surprised
by B.~L.~Ginzburg; in the second aspect (the knowledge of numerical data),
by Ya.~B.~Zeldovich, who, even speaking over phone,
could mentally calculate whether or not a given theorem corresponds to physics.

If a mathematician does not have any physical intuition
(for example, just as the author\footnote{In particular,
my model of the Chernobyl pile-up cooling turned out to be erroneous
(see~\cite{Masl_Molotkov}).}),
then he can be severely disappointed
considering some well-known physical law as an axiom and obtaining
an answer contradicting the experimental data.
Therefore, prior to presenting mathematically rigorous results,
which correspond to the unsolved problems of chemistry
(and even of alchemy, because the term ``fluids'' was introduced
by alchemists\footnote{The alchemists were not far from truth, for example,
addition of 1\% of methanol to a carbon dioxide fluid
results in that $CO_2$ on the critical isotherm behaves as the 100\% methanol
in several extraction processes. In technology, the supercritical water vapor
is used in turbines in all thermal and nuclear power stations.}),
we show that some well-known, even ``eternal'' physical and mathematical ``verities''
turn out to be significantly refined in the case of rigorous mathematical verification.

\begin{remark} 
To distinguish the classical theory in its modern understanding
from the quantum theory, it is necessary somewhat to change
the physicist's ideology that the classical theory is the theory
that existed in the 19th century before the appearance of the quantum
theory. But, in fact, the classical theory is the theory obtained
from the quantum theory in the limit as $h\to 0$.

So Feynman was right stating that spin is a phenomenon of classical mechanics.
Indeed, this is the case in rigorous passing to the limit
from quantum to classical mechanics.
Similarly, the ray polarization does not disappear
as the frequency increases and hence is a property
of geometric rather than wave optics, as was used to think
because the light polarization was discovered
as a result of discovering the wave optics.

We consider the ``Lifshits well'', i.e.,
the one-dimensional Schr\"odinger equation with potential
symmetric with respect to the origin and having two wells.
Its eigenfunctions are symmetric or antisymmetric
with respect to the origin.
As $h\to 0$, this symmetry remains unchanged,
and since the squared modulus of the eigenfunction corresponds
to the probability that a particle stays in the wells,
it follows that, in the limit as $h\to0$, i.e.,
``in the classics'', for energies less than the barrier height,
the particle occurs between the wells, in the two wells at once,
although the classical particle cannot penetrate through the barrier.
Nevertheless, this simple example shows the variation in the ideology
of the ``classical theory''.

To understand this paradox, it is necessary to take into account
that the symmetry must be very precise and the state stationarity
means that this state appears in the limit of ``infinitely large'' time.
As we show below, the same concerns the Bose-distribution
in the classical theory of gases.

Now we show how the existence of an additional parameter
changes the representation in the classical limit.
We dwell upon the notion, which is called ``collective oscillations''
in classical physics and ''quaseparticles'' in quantum physics.
In classical physics,
this is the Vlasov equation of self-consistent (or mean) field,
and in quantum physics, this is the Hartree (or Hartree--Fock) equation.

So we want to note the following point that seems to be paradoxical.
The solutions of the equation in variations for the Vlasov equation
\textit{do not coincide} with the classical limit
for the equations in variations for the mean-field equation
in the quantum theory. This is because of the fact that the variations
are related to another small parameter, namely, the variation parameter.
And this is already the field of nonstandard analysis.

N.~N.~Bogolyubov, for example, in \cite{Bogol}, studied
the problem without an external field,
and the asymptotics thus obtained coincides
with the semiclassical asymptotics in an external field
\cite{QusiPart_1},~\cite{QusiPart_2}.
Moreover, this is in fact the classical limit,
because the parameter $h^2$ in this Bogolyubov's paper
can be compensated by a large parameter, namely, the wave number~$k$.
This becomes especially obvious if the interaction is assumed to be zero.
Then an ideal gas is obtained, which in this case can be considered
as a classical gas.
Just therefore, the author proved that the classical fluids
in nanotubes have superfluidity
(see \cite{RJ_SuperFL},~\cite{TMF_153_3}),
which was confirmed by a series of experiments
(see the references in~\cite{RJ_SuperFL}).

Confusion is due to the fact that the constant~$h$
has dimension of action, and, rather often, in the classical limit,
in order to keep the dimensions, it participates even in the
Maxwell distribution \cite{Landau}.
For example, the Thomas--Fermi equations
(including the temperature equations) are classical equations
from the above point of view, but the fact that they contain
the constant~$h$ leads to confusion.
Therefore, the Van der Waals law of corresponding states is very important,
it allows us to consider the reduced temperatures
$T_\text{r}=T/T_\text{cr}$ and pressures $P_\text{r}=P/P_\text{cr}$
as dimensionless quantities.
\end{remark}

\begin{remark}
Now we consider the $N$-particle Gibbs distribution
\begin{equation}\label{2}
\mathcal{P} = e^{-\frac{H(q,p)}{kT}}, \qquad H(q,p)=\frac{p^2}{2m}
+ U(q), \qquad q\in R^{3N}, \qquad p\in R^{3N}
\end{equation}
where $H(q,p)$ is the Hamiltonian.
It is assumed that the level surfaces $H(p,q)=\text{const}$
are simply connected.

The theorem about this distribution was in fact proved in~\cite{TMF_155_2}.
It must be considered in the sense of Kolmogorov complexity.
This distribution is the distribution over the number
of different experiments whose number~$L$ is independent of~$N$,
over systems of $N$ particles at the same temperature
(the average distribution over the number of experiments),
and it is the distribution over the energy surfaces $
H(p,q)=\text{const}$, $p\in R^{3N}$, $q\in R^{3N}$:
$$
E_1\leq H(p,q) \leq E_2,
$$
where $E_1$ and $E_2$ are less than some average energy~$E$,
$\Delta E=E_1-E_2\ll E$, and the phase volume is $\Delta\omega=\int\,dp\,dq$,
$E_1\leq H(p,q)\leq E_2$.

We divide the phase space $(p,q) \in R^{6N}$ into finitely many domains
\begin{equation}\label{6a}
E_l\leq H(p,q)\leq E_{l+1},
\end{equation}
where $l=0,\dots, s-1$, $E_0=0$, $E_s=E$, $p\in R^{3N}$, $q\in R^{3N}$,
and, correspondingly, the phase space $R^{2NL}$ has coordinates
$p_1, q_1, p_2,q_2, \dots, p_{N}, q_{N}$.
We perform ordered sampling with return $L_l$
from the partition of domains in the space $R^{2NL}$
into the ``box'' $E_l\leq H(p,q)\leq E_{l+1}$,
under the condition that
\begin{equation}\label{7}
\sum_{i=1}^{L}\frac{1}{\Delta\omega}\int_{L_i\leq  H(p,q)\leq
L_{i+1}} H(p,q)\,dp\,dq.
\end{equation}

From the physical viewpoint, the ordered sampling means that~$L$
distinguishable $3N$-dimensional particles are considered.
Let $\rho^\Delta_{E_l}$ be the number of $3N$-dimensional ``particles''
in the energy interval $E_l\leq H(p,q)\leq E_{l+\Delta}$
divided by~$\Delta\Omega$.

Assume that the above conditions on the function $H(p,q)$
are satisfied.
We determined $L$ from the condition\footnote{If the temperature $kT$
is considered as average over the the number of particles,
then $b=\frac{1}{kNT}$. Here $kNT$ is the average energy over the number
of experiments.}
\begin{equation}\label{8}
\frac{1}{\Delta\omega}\int_0^\infty  e^{-b H(p,q)}\,dp\,dq =L
\qquad b=\frac{1}{kT}, \qquad p,q \in R^{6N},
\end{equation}
where $k$ is the Boltzmann constant. We determine $E$ in~\eqref{7} as
$\int_0^\infty H(p,q)e^{-bH(p,q)}\,dp\,dq$.
Then we have the following theorem.

\begin{theorem}\label{theor2} The following relation holds:
\begin{equation}\label{9}
\mathcal{P}\biggl(\bigg| L
\rho^\Delta_{E_l}-\frac{1}{\Delta\omega}\int_{E_{l} \leq
H(p,q)\leq E_{l+1}} e^{-b H(p,q)}\,dp \,dq \bigg| \geq \sqrt{L\ln
L} \biggr) \leq L^{-m},
\end{equation}
where $m$ is arbitrarily integer.
\end{theorem}

Here the probability $\mathcal{P}$ is the Lebesgue measure of the phase
volume in parentheses in~\eqref{9}
with respect to the entire phase volume $R^{6NL}$
bounded by~\eqref{7}.

Thus, the Gibbs distribution is not a distribution
over momenta and coordinates, but is a distribution over the energy levels.

If the domain~\eqref{6a} is multiply connected,
then the problem can be solved only by semiclassical transitions.
If $N=1$, then we formally obtain the Maxwell--Boltzmann distribution.
But the latter is considered, as a rule, as the average distribution
for $N$ particles. We discuss this treatment in Remark~5.

Boltzmann obtained his distribution integrating this distribution
over~$p$.  But this can be done under the following conditions only.

For a great many particles, their dependence on a common potential field
must be assumed to be slowly varying as a rule.
Indeed, if the particles are in a volume~$V$
(or in an area~$S$), then the thermodynamical asymptotics requires
that the volume~$V$ (the area~$S$) tend to infinity.
But this means that the potential field varies very slowly, and it must
be assumed that the function of the coordinates is of the form
$$
H(p, \frac{q}{\sqrt[3]{V}}).
$$
Otherwise, we have no rights to pass from the Gibbs distribution
from the Maxwell--Boltzmann distribution $e^{-\beta(p^2+U(q))}$
(integrating it over the momenta)
to the Boltzmann distribution of the form
\begin{equation}\label{Bol}
e^{-\beta U(x)},
\end{equation}
since the Maxwell--Boltzmann distribution is not a distribution
of the density of the number of particles with respect to the momenta
and coordinates. This distribution can give only the number of particles
between energy levels of the form
$$
\frac{p^2}{2m}+ U(x) = \text{const}.
$$
But if $U(q)$ is of the form
$$
U(q) = F(\frac{q}{\sqrt[3]{V}}),
$$
then this integration can be done in the thermodynamical asymptotics
as $V\to\infty$.

One should not forget here that the Boltzmann distribution
is also far from being a distribution with respect to the coordinates~$q$,
and that it is a distribution with respect to level surfaces of the function
$U(q)$ only.
\end{remark}

\begin{remark}
There is a mathematical error in the very definition of
thermodynamical limit as the limit as $N\to\infty$ and $V\to\infty$
such that $N/V=\rho$, where $\rho$ (the density) is finite.

Let the pressure be $P\sim 1/23\ \text{atm}$.
This pressure does not lead to a too rarified gas
(and is admitted by the Knudsen criterion).

Obviously, the above definition is incorrect because the number of
particles is bounded by Avogadro's number.
As $N\to\infty$, we have $\ln\ N\to\infty$ as well.
But $\ln\ N\sim 23$ and $\ln\ N$ is not large under the above pressure.
It would be correct to pose the problem, as was said above,
neglecting the inverse quantity $(\ln^2N)^{-1}$.
In this case, a mathematically rigorous asymptotics
in statistical physics would be obtained.
\end{remark}

\begin{remark}
The transfer of the Bose distribution for photons to gas
in the form of the Bose--Einstein distribution is incorrect.
Since $\sum N_i=N$, where $N_i$ stands for the number of particles
at the energy level $\varepsilon_i$, cannot vary up to infinity
($N_i\leq N$), it follows that
the chemical potential $\mu$ can take positive values,
which must not be neglected \cite{ArXiv_Ec}. This restriction means
that parastatistics must be considered.
\end{remark}

\begin{remark}
The most essential is the note that the distribution
given in Remark~4 with the property that $N$ and $\ln\,N$ are finite
permits preserving the parastatistics of the Bose--Einstein-type distribution as $h\to 0$,
and just this property will be used in what follows.
The fact that it holds in the classical limit
is especially unusual for the physicists (cf. Remark~1.).

Therefore, in addition to references to exact theorems,
I also present several other arguments. For example,
the physicists completely accepted the financial considerations
presented in~\cite{QuantEcon}.

``A general important property of money bills is that
the change of one money bill by another of the same denomination
does not play any role.
This property, which can be called ``money do not smell'',
permits uniquely determining the formula for nonlinear addition.
We assume that we want to deposit two copecks into two equivalent banks.
How many possibilities do we have? We can deposit both copecks
into the first bank, or deposit both copecks into the second bank,
or deposit per one copeck in each of the banks.
Thus, we have three possibilities.
But if we want to deposit two diamonds, then we have four possibilities,
because we can interchange the diamonds.
At the same time, it does not make any sense to interchange the copecks.
The identity property of money bills of the same denomination
permits changing the number of versions.
This statistics, just as the statistics of identical Bose-particles
is called the quantum statistics (Bose-statistics),
and the statistics of diamonds (provided that they are not absolutely
identical) is called the classical statistics.
But as we see, the Bose-statistics can be applied to money, more precisely,
to money bills.''\cite{QuantEcon}, p.~5.
And this is despite the fact that the copecks
may have different years of their issue
and the money bills have individual numbers.
Objectively, they are different, but for our problems,
this difference is not essential, because only the quantity
of money bills is important for us.

Further, the physicists did not protest, when Schoenberg used
the creation and annihilation operators in classical mechanics~\cite{Masl_Shved}.
And this is possible only when considering the Fock space of
identical particles.

We consider one more aspect of this problem in more detail,
because the molecular dynamics of classical particles
is nowadays commonly used.

In the well-known textbook \cite{Landau_Kv_Mech} by Landau and Lifshits,
the authors explain the identity principle for particles as follows:
``In classical mechanics, identical particles (such as electrons)
do not lose their `identity' despite the identity of their physical
properties. Namely, we can visualize the particles constituting
a given physical system at some instant of time as `numbered'
and then observe the motion of each of them along its trajectory;
hence, at any instant of time, the particles can be identified. ...
In quantum mechanics, it is not possible, in principle, to observe each
of the identical particles and thus distinguish them.
We can say that, in quantum mechanics, identical particles
completely lose their `identity' ''~(p.~252).

There are similar explanations of the identity principle for particles
in other textbooks as well.

But, as a matter of fact, if the initial data for the Cauchy problem
does not possess a symmetry property, then the situation in
quantum mechanics does not differ from that in classical mechanics.

Indeed, suppose that the Hamiltonian function
$$
H(x_1,p_1; x_2,p_2; \dots; x_n,p_n)
$$
is symmetric with respect to permutation of $(x_i,p_i)$ and
$(x_j,p_j)$,   \, $i= 1,2,\dots, n; \, j= 1,2,\dots, n$.
Let $\widehat{H} = H(x_1,\widehat{p}_1; \dots; x_n,
\widehat{p}_n)$ denote the self-adjoint operator in the space
$L_2(R^{3n})$ corresponding to the Hamiltonian
$H(x_1,p_1; x_2,p_2; \dots; x_n,p_n)$
(Weyl of Jordan quantized).

Consider the corresponding Schr\"odinger equation
\begin{equation}\label{5_0}
ih\frac{\partial\Psi(x)}{\partial t}= \widehat{H}\Psi(x), \qquad
\Psi(x)\in L_2(R^{3n}),
\end{equation}
satisfying the initial conditions
\begin{equation}\label{5_1}
\Psi|_{t=0} =\Psi_1(x_1) \Psi_2(x_2) \dots\Psi_n(x_n) \exp \frac
ih\sum_{i=1}^n p_jx_j,
\end{equation}
where $\Psi_i(x) \neq \Psi_j(x), \, p_i \neq p_j$,
and we can assume that
$\Psi_i(x)= f(x-x^{(i)})$, where $x^{(i)} \neq x^{(j)}$,
and $f(x)$ is a bell-shaped function vanishing outside a neighborhood
of the point $x^{(i)}$, and the neighborhoods are small
so that these functions do not intersect.
Returning in time to the initial point, we can number
all the bell-shaped functions.

But in the projection on the real space $R^3$
containing all~$n$ particles, the experimenter cannot distinguish
them without taking into account the full deterministic process
with respect to time from zero to the given~$t$.

Similarly, in classical mechanics, of two point particles
have intersected and the experimenter has no knowledge of their velocities
(instant photo), then he also cannot distinguish them.
He must know their original velocities, i.e., use slow-motion filming.
This means that he has to look into their ``past''.

But if the experimenter must determine which of the original particles
with a prescribed velocity arrived at the given point,
then he must observe the whole process, up to the point $t=0$.

Finally, let us consider the particle identity philosophically.

In statistical calculations of the number of inhabitants in
a town, the permutation between a child and an old man does not change
the total number of inhabitants. Hence, from the point of view of the
statistics of the given calculation, they are indistinguishable.
From the point of view of the experimenters who observes
the molecules of a homogeneous gas using an atomic microscope,
they are indistinguishable. He counts the number of molecules
(monomers) and, for example, of dimers in a given volume.
Dimers constitute 7\% in the total volume of gas (according to
M.~H.~Kalos). This means that the experimenter does not distinguish
individual monomers, as well as dimers, from one another and counts
their separate numbers. His answer does not depend on the method
of numbering the molecules.

These obvious considerations are given for the benefit of those
physicists who relate the fact that quantum particles are
indistinguishable with the impossibility of knowing the world.
I do not intend to argue with this philosophical fact, but wish
to dwell only on mathematics and statistics and distributions
related to the number of objects.

Let us turn to the Boltzmann statistics, as is described
in ``Mathematical Encyclopedia'' \cite{Math_En}.
In his article on the Boltzmann statistics, D.~Zubarev, the well-known
specialist in mathematical physics and the closest disciple of
N.~N.~Bogolyubov, writes that, in Boltzmann statistics,
``particles ... are distinguishable''.
However, a few lines below, Zubarev states that ``In the calculation
of statistical weight, one takes into account the fact that the permutation
of identical particles does not change the state, and hence the phase
volume must be decreases by $N!$ times''.

Of course, it is impossible to simultaneously take into account both remarks.
But they are both needed to solve the Gibbs paradox.
As I pointed out on numerous occasions, from the mathematical point of view,
the Gibbs paradox is a counterexample to the Maxwell--Boltzmann distribution
regarded as a statistical distribution for a gas in which
the molecules cannot be turned back, but not a counterexample
to a dynamical distribution~\cite{Kozlov}. As for the famous discussion
between Boltzmann and some mathematicians~\cite{Boltzman},
of course, if the particles of the gas are distinguishable
and can be numbered, then they can also be mentally turned back
and returned to a state close to their initial state,
as the Poincare theorem states.

As an example we consider the last source of errors,
because the computer has a finite memory.
How is the number of particles related to the computer error
and the computer's ability to recover their initial data?

By way of example, we consider
a set of unnumbered billiard balls of unit mass and the same color.

First, consider one billiard ball and launch it from some (arbitrary)
point with velocity not exceeding a certain sufficiently large value~$v$,
i.e., with energy not exceeding~$v^2/2$.
However, since the computer has certain accuracy, it follows that
the energy of the ball will take a finite integer number, $s$,
of values in the interval of energies $[0,v^2/2]$:
$\lambda_i=iE_0$, $i=1,\dots, s$, where $E_0$ corresponds to this accuracy.

Thus, we obtain a spectrum of energy values which can be regarded
as a self-adjoint diagonal matrix of order~$s$,where $s\gg 1$.
By assigning such a discrete set of energies to a ball,
we obtain the wave--particle correspondence in classical mechanics,
because the resulting matrix is unitary equivalent
to any operator $\widehat{L}$ with such a spectrum in a Hilbert space~$H$.

How many balls must be launched so that the computer will be unable to
determine their initial data?

The spectrum corresponding to~$N$ balls is obtained by considering
the tensor product of~$N$ Hilbert spaces:
$\widehat{L}_N =\widehat{L}\times\widehat{L}\times \dots \times\widehat{L}_{(N
\text{times})}$.

The eigenvalues of this operator have the form
$$
{\cal{E}}=\sum_{i=1}^s N_i\lambda_i.
$$
If we only consider the eigenvalues symmetric with respect to the
permutation of the particles of this operator,
which corresponds to the identity of the balls, then
the eigenvalue ${\cal{E}}=\sum_{i=1}^s N_i\lambda_i$
is of multiplicity equal to the number of all possible variants
of the solution of the problem
\begin{equation}\label{5_3}
{\cal{E}}=\sum_{i=1}^s N_i\lambda_i; \quad   \sum_{i=1}^s N_i  =N.
\end{equation}

Since the initial set of energies is ``without preferences,''
i.e., is in general position, then all the multiplicities
corresponding to~\eqref{5_3} are equiprobable.
The computer calculation time is related to the computer
accuracy~$E_0$ with respect to energy. The problem is how to determined~$N$
for which, at a particular instant of time, the computer cannot recover
the initial data in view of the inaccuracy of the classical pattern
or that of the quantum mechanical pattern (which is more accurate,
but more cumbersome); the latter pattern, in turn, is not accurately
described by an interaction of Lennard-Jones type.

Thus, we can draw the following conclusion. The initial data
in the classical and quantum mechanical problems are forgotten
because of external noise. As a result, the problem is reduced to
the distribution $\{N_i\}$ \eqref{5_3}.
In this problem, we assume {\it a priori}
that the initial data are forgotten and, therefore, so is the numbering of
classical particles. Although the textbook~\cite{Landau_Kv_Mech}
erroneously interprets the difference between the quantum mechanical pattern
and the classical one, it, nevertheless, gives a valid interpretation
of the numbering of identical balls. Therefore, we can take symmetric
eigenfunctions for~$\widehat{L}_N$.

Therefore, it only remains to obtain the distribution
of the number of particles~$N_i$ using relations~\eqref{5_3}.
If $s\gg N$, then~\eqref{5_3} can be expressed as
\begin{equation}\label{5_4}
E_0\sum_{i=1}^\infty i N_i={\cal{E}}; \qquad   \sum_{i=1}^\infty
N_i  =N,
\end{equation}
these relations coincide with those in the classical number-theoretic problem
under the condition that ${\cal{E}}/E_0$ is integer, which, of course,
is of no importance in the asymptotics as $s\to\infty$ and $N\to\infty$.

Thus, since the noise component has prevented us to recover the initial data
and the number of particles $N$ is preserved as well as
the total energy~${\cal{E}}$, without given any preference\footnote{For
a market model ``without preference,'' see~\cite{QuantEcon}}
to any versions from~\eqref{5_3},
we assume all the variants satisfying the relation
\begin{equation}\label{5_5}
\sum i N_i\leq \frac {\cal{E}}{E_0}, \quad \sum N_i=N,
\end{equation}
to be equiprobable.

In this approach, we can also take into account the collision of billiard balls,
the initial energy of all balls can only decrease due to friction and the
passage of kinetic energy into thermal energy during collisions.
The numb er of balls will remain the same and the total energy will not exceed
the initial energy~${\cal{E}}$.

This approach can also be applied to turbulence if we consider
such a significant invariant as the Kolmogorov spectrum.

Note another highly important difference between the statistical and
dynamical approaches. If the experimenter presents some data, then, as a rule,
this means that he has carried out many experiments
and presented the data consistent with all the experiments.
In addition, these data can be verified by other experimenters and only
after that they will be universally accepted.
This means that all the variants of relations~\eqref{5_5} are,
in a sense, equiprobable.
If inequality~\eqref{5_5} is replaced by the corresponding equality,
then we obtain the well-known microcanonical distribution.
But as shown in~\cite{MN_78_6}, the final results for the inequality
and the equality coincide. Therefore, we have finally derived
the well-known microcanonical distribution for the number of particles,
whose permutation does not affect the counting result.

Thus the notion of different (distinguishable) particles
is incorrect if one speaks about their density. From the point of view
of density of the set of objects under consideration, these objects are
indistinguishable, namely, any transportation of these objects does not
influence the density.
Therefore, one must not apply the Boltzmann distribution
in thermodynamics. Since~$N$ is a large number and the value $\ln\,N$
is not large, one must use parastatistics
(see \cite{ArXiv_Ec},~\cite{RJMP_2009_2},~\cite{RJ_16_1}
and formula~\eqref{eq3} below).
\end{remark}

So we arrived at some inconsistencies in the main laws of
thermodynamics and statistical physics. One of the most important
distinctions between them is that there is a positive chemical potential
in the distribution for the ``Bose gas'' of finitely many particles~$N$.
Of course, this is related to the famous notion of Bose condensate.
As was shown in~\cite{Landau}, there exists a degeneracy temperature
such that for $T=T_{\text{deg}}$ the number of particles is maximal,
$N_{\max}$, and if $N>N_{\max}$, then $N-N_{\max}$ of particles
get into the Bose condensate with energy zero (or minimal).
We present an example of such a condensate in number theory.

For this, we consider Koroviev--Fagot's trick,
described by M.~A.~Bulgakov in the novel ``Master and Margarita'',
which was already discussed in our preceding
papers~\cite{RJ_16_1, MN_85_3}.
In a variety show, he scattered $n$ money bills
of 1 tchervonets denomination among $k$ spectators.
We assume that all versions of scattering of $n$ money bills
over~$k$ spectators are equiprobable (the chaotic condition).
This is equivalent to decomposition of an integer~$n$ into~$k$ terms.
The logarithm of the number of possible versions $p_k(n)$
is the entropy. We consider two examples related to Fagot' trick.

Let $n$ money bills be distributed over $k_0$ spectators, where $k_0$
is such that
$$
p_{k_0}(n)= \text{sup}_k p_k(n).
$$
(Clearly, for $k=1$ and $k=n$, the number of versions is equal to~$1$,
and hence such a supremum exists.) Then, with a larger probability,
$k_0$ spectators obtain a quantity of money bills.
If for $k>k_0$, the number of versions and hence the entropy decrease,
then the entropy is maximal when the remaining spectators do not get
any bill, i.e., they get into the Bose condensate\footnote{As follows
from relation~\eqref{5_5} known in this problem, this situation
corresponds to the two-dimensional case and demonstrated
the Bose condensate in the two-dimensional case with Remark~3
taken into account.}.

Now we assume that the number of spectators is $2k_0$.
Then, roughly speaking, $k_0$ (i.e., a half) spectators
obtain several money bills, and $k_0$ spectators do not obtain anything.
Since $k_0$ spectators with a larger probability
(see~\cite{ArXiv_Ec}) obtain zero of money bills,
the number of versions of the distribution of money bills
over spectators is also equal to $p_{k_0}(n)$.

We couple the spectators under the following condition:
if somebody in a pair does not obtain any money bill,
then the sum obtained by the other is divided, for example, in half.
Therefore, we now have $k_0$ pairs,
and the number of all versions is equal to the number of versions
in the first example.

Therefore, the Kolmogorov complexity (entropy) in both examples
is the same and equal to $\log_2p_{k_0}(n)$, where $k_0$ is the number
of pairs. Thus either case is equiprobable, i.e.
it there were $2k_0$ spectators, then $k_0$ of them would not get anything
and if the spectators coupled, then all the pairs would obtain
some money bills, but the number of versions of the distribution
still remains equal to $p_{k_0}(n)$,
and none of the examples can be preferred.
Both of the realizations is equiprobable.
In other words, it may be that the spectators coupled,
and it may be that $k_0$ spectators remain without any money bills.

If we speak about the social economic efficiency,
then it would be better if the spectators coupled.

Thus, it is a priori unknown what is the results of the phase transition
of an ideal Bose gas into condensate:
pairs of particles or, as was previously assumed in physics,
particles at the ground level.
If we consider the repulsing potential,
as N.~N.~Bogolyubov did in his famous work in 1947,
then this problem remains unsolved.
But if we consider the Lennard--Jones potential, then,
as the author showed in~\cite{ArxivGas}, there are arising pairs.
In this case, one significantly uses the
Pontryagin--Andronov--Vitt theorem~\cite{Pontr}, which implies
that the time of stay of a particle inside the barrier,
which exists in the case of the Lennard--Jones potential,
depends on the noise level or the gas temperature.

It is clear that the transition into the condensate state
must occur with a larger probability, which a mathematician
must estimate. These estimates were first performed in~\cite{ArXiv_Ec}.

\section{Problem of fluids}

If some particles form pairs (dimers), then the free particles
(monomers) occur in the fractal situation.
This fractal dimension can in general be counted by
the Hausdorff formula.

As is known, as the temperature increases,
the number of dimers decreases, and hence the fractal dimension increases.
And this means that one more pair of quantities must be introduced
in thermodynamics: the extensive (dimension) quantity
and the intensive quantity corresponding to it,
which we call the ``press''.
This can be done rather easily, because the thermodynamical
potential $\Omega$ is known for parastatistics of any dimension.
Its derivative with respect to the dimension is just the desired
intensive quantity~\cite{Landau}.

\begin{figure}
\begin{center}
\includegraphics[width=5cm]{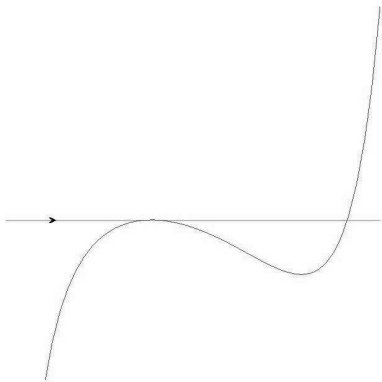}
\caption{The trap for a fictitious particle in the system
of the center of mass (SCM).} \label{fig1}
\end{center}
\end{figure}

As for the problem shown in Fig.~\ref{fig1},
it is clear that, for $E\leq E_{\min}$,
there exist only monomers, and between $E_{\min}$ and $E_{\max}$,
i.e., between the barrier bottom and height, there exist both monomers
and dimers. Hence the fractal dimension between $E_{\min}$ and $E_{\max}$
is less than~$3$.

Let us consider the following natural variation of the notion of
Bose condensate for the classical parastatistical gas.
Since the parastatistical condensate results in the appearance
of pairs (dimers), the number of free degrees of freedom increases,
and hence one can speak not about the energy disappearance
in the parastatistical condensate but about the energy conservation
by means of its redistribution.
But if dimers, i.e., molecule associations, appear,
then rotational and oscillatory components are added.
This decreases the total energy of monomers, and hence
$N_{\max}$ also decreases, which started an avalanche-like process
until (with a larger probability)
the translational energy of monomers disappears.
And hence there is a phase transition in the energy redistribution
and, naturally, a phase transition in the fractal dimension,
which thus decreases.

We must not seek a phase transition of such a type in the domain,
where the gas--liquid transition occurs according to
the Van der Waals--Maxwell scheme, the scheme accepted
by both the specialists in thermodynamics
and the specialists in molecular physics of the old school.
But the specialists in modern mathematical physics prefer to use
the theory of percolation and renorm-group,
as well as the possibilities due to computer computations.

The specialist in the theory of nucleation, especially experimenters,
try to invent new theories and say that the classical theories
are of no use anymore\footnote{Cf. the opinion of the famous experimenters
R.~Strey and H.~Reiss: ``Present-day experiments
in the homogeneous formation of droplets in argon
display a dramatic discrepancy between the experimental data and
the theoretical prediction based on the widely used old classical
theory''.}.

Both in the theory of percolation and in the theory of nucleation,
as well as in the Van der Waals--Maxwell theory,
an important role is played by the surface tension, treated
as an additional energy.
If this additional component is not considered,
then the just described purely mathematical approach
and the phase transition must be sought in the domain of
critical temperature, i.e., in the domain, where the surface
tension disappears. This structure,
in which gas and liquid are indistinguishable, is called fluids
from the time of alchemists.

Just as in Remark~2, we can write the microcanonical distribution,
but under the condition that permutations in $N_i$ are forbidden.
As a result, we obtain a parastatistical distribution
that is very close to the Bose distribution.

As was already noted, difficulties aries when the domains~\eqref{6a}
are multiply connected. Moreover, a special case is the case
of continuous spectrum.

As a rule, in physics in this case, the correct conditions
of the Sommerfeld radiation are posed, which turn the problem
into a not self-adjoint problem.
If we go out into the complex domain through the cut,
then it is reduced to the problem of finding Regge-type poles
for a self-adjoint problem.

The mathematicians considered, as an example,
a potential on a finite interval in the half-plane
for the one-dimensional wave equation with a finite well
and with the radiation conditions on the boundary of the finite domain
(the domain where the potential takes the values $[0, l]$).

Let $l_1$ be the value of $x$ on the barrier maximum.
But the ``eigenfunctions'' corresponding to the Regge poles
(the poled of the resolvent continued through the cut into the complex domain)
increase exponentially as $x\to \infty$.
But the Regge ``eigenfunctions'' for which the real parts of the poles
are close to the quasilevels of the well
are sufficiently large in the domain of the well.

We prove that the solution of the Cauchy problem for the wave equation
whose initial conditions are consistent with the radiation conditions
(i.e., the derivative with respect to~$t$ depends on the initial function
for  $t=0$) tends to zero at each point~$x$ of the support
of the finite potential.

On the one hand, the passage to the semiclassics as $h\to 0$
simplifies the problem, on the other hand, there are two
interacting parameters $h\to 0$ and $t\to\infty$.

In the problem under study, there is addition energy
equal to the average energy $E_{\text{av}}=kT$
(where $k$ is the Boltzmann constant and $T$ is the temperature)
multiplied by the number of particles~$N$ (another large parameter).

Thus, we must compare the quantity $h/t$ with $NE_{\text{av}}$.
In fact, there is one more dissipation in gas, namely,
the viscosity, which also cannot be neglected in mathematical
computations and which ensures that the initial problem
is not self-adjoint (the parameter corresponding to friction).
The problem of determining the limit situation
becomes significantly simpler in the language of nonstandard analysis.
The rigorous theorems even in the case of this simplification
are very c cumbersome. We present the main ideas.

In the semiclassical limit, one can assume that
the well up to the barrier maximum and the domain
from this point to the boundary condition are in no way related to each other.
Only at the point $E$ equal to the barrier maximum in infinite time,
the classical particle can ``creep'' over the barrier (see Fig.~\ref{fig1}).

The mathematicians consider the one-dimensional wave equation
$$
h^2\frac{\partial^2\Psi}{\partial t^2}=
h^2\frac{\partial^2\Psi}{\partial x^2}+ u(x)\Psi
$$
in the half-space $x\geq 0$, $u(x)=0$ for $x>l$.
The potential $u(x)$ has a smooth barrier with maximum at the point
$x=l_1$. The stationary problem has the form
$$
h^2\frac{d^2\varphi}{d x^2}+u(x)\varphi= k^2\varphi.
$$
The radiation conditions outside the interval $[0,l]$
have the form $ \text{const}\,e^{(-ikx)/h}$.
The initial condition is
$$
\Psi|_{t=0}=\Psi_0(x)  \in L_2 [0,l],
$$
and $\Psi'|_{t=0}$ is chosen so that it correspond to the travelling
wave outside the interval as $x\to\infty$.
The residues at the Regge poles form a basis on this interval $[0,l]$.

Let $\Xi$ be the linear span corresponding for $h\ll 1$
to all quasi-eigenfunctions whose eigenvalues are less than
$E_{\max}$, and $P_{E_{\max}}$ is the projection on this subspace.

Let the initial function be equal to $\Psi(x) \in L_2[0,l]$.
Then its part corresponding to the well is equal to
$P_{E_{\max}}\Psi_0(x)$, and this part as $t\to\infty$
tends to zero in the interior of the domain of the well.

Now we consider the contrary problem, i.e., the ``irradiation'' problem.
In other words, we pose conditions that are complex conjugate
to the Sommerfeld conditions. This problem does not always has solutions.
The well is being filled until
$$
P_{E_{\max}}\Psi(x,t)\theta(x-l_1)=\Psi(x,t)\theta(x-l_1).
$$
If
$$
P_{E_{\max}}\Psi(x,t)\theta(x-l_1) \neq \Psi(x,t)\theta(x-l_1),
$$
then the solutions do not exist, because the part
$$
\big(1-P_{E_{\max}} \big) \big[\Psi(x,t)\theta(x-l_1)\big]
$$
reflects from the point $x=0$,
and the ``irradiation'' condition is not satisfied.

This phenomenon is similar to the following one.
Assume that a negatively charged body is ``irradiated'' by positive ions.
Then after a certain number of ions sticks to the body
(fills the well) and neutralizes it, the other freely flying ions reflect
from the body and the ``irradiation'' process stops.
The conditions dual to the Sommerfeld condition are not satisfied any more,
because the radiation process originates simultaneously.
In our case, the role of attracting charges is played by Regge poles.

The Sommerfeld conditions are a specific case of the blackbody condition.
This problem was posed even by Frank and von Mises in their survey
as one of fundamental problems. The Feynman integral
does not contain any measure, because it is impossible to pose
the boundary conditions for the Feynman tube:
the Feynman tube boundary is a black body.
By definition, if the trajectory enters this boundary,
it should not be considered any more,

In the Wiener integral of tunes, this is also a black body,
but compared with the wave problem,
the black body in the diffusion problem is understood as
a situation in which the particles stick to the wall,
and this means the zero boundary conditions for the heat
conduction equation.
After the transition to $p$-representations, as was shown in
our work with A.~M.~Chebotarev~\cite{Masl_Cheb},
the zero boundary conditions give the Feynman tube
in the $p$-representation, and it is possible to determine
the measure for the finite potential.

The temporary energy capture  in the laser illumination
and its subsequent transformation into a directed beam
also characterize the absorption-type trap of ``irradiation''
that further turns into radiation.
Probably, the problem of black holes in astronomy
is a problem of the same type.
In the one-dimensional case, on a finite interval,
the modern mathematical apparats can solve this problem completely.
In this case, if $E_{\text{total}} = E_{\text{av}}N$,
where $N$ is the number of particles,
then with a large probability) one can determine
the number of particles inside the well and the number of particles
outside it. By the way, this also readily follows from the concept
of microcanonical distribution.

Indeed, the multiplicity
(or the ``cell'' as it is called in~\cite{Landau})
of the eigenvalue $E_{\text{av}}$ is different
inside and outside the well:
it depends on the density of eigenvalues as $h\to 0$
on a small interval $(E_{\text{av}}-\delta,E_{\text{av}})$.
Obviously, this quantity is proportional to the width
of each well on the interval $[0,l]$, i.e.,
to the length of $[0,l_1]$ and $[l_1,l]$.

Assume that $N_1$ is the number of particles in the well $[0,l_1]$,
$N_2$ is the number of particles in the well $[l_1,l]$,
$d_1$ is the length of the interval $[0,l_1]$,
and $d_2$ is the length of the interval $[l_1,l]$.
Then $(d_2/(d_1+d_2))N$ of particles will reflect.

If $d_1>d_2$, then the well--trap is not overfilled,
and the irradiated particles remain in the well--trap.
But if $d_2>d_1$, then $(1-d_2/(d_1+d_2))N$ of particle
fall out of the trap and reflect.

In numerous papers, the author has rigorously proved that
there exists a parastatistical condensate similar to the Bose condensate
in the one-dimensional and two-dimensional cases as well as in the case
of fractal dimension greater than zero.
This holds not only for identical objects (particles),
but for objects (particles) that are assumed indistinguishable,
such as those in the calculation of the ``number'' of teenagers,
the ``number'' of women, the ``number'' of men,
the ``number'' of money bills of one denomination,the ``number'' of
cites with population of over 1000 people, etc.
This same applies when we speak about density.
In the last case, the difference with thermodynamics is in only that
there can be no finite thermodynamic limit
(as stated in~\cite{MZ_86_3}), i.e.,
the ratio $N/V$, where $N$ is the number of objects,
and~$V$ is their volume, area, or length, may tend to infinity.

In this case, however, other small parameters that ``neutralize''
this convergence to infinity may exist;
for example, as show in the previous paper~\cite{MZ_86_3},
these may be nanodimensions (nanotubes).

In addition, the author proved that the parastatistical condensate,
which is close to the Bose--Einstein condensate, does not necessarily
lead to the accumulation of particles at the lowest energy level
(roughly speaking, when they ``stop''), but can lead to the formation
of pairs of particles (dimes) and clusters and to the conservation
of energy, i.e., passage of translational energy into rotational
oscillatory energy.

In the present paper, we show that this point of view for the thermodynamics
of imperfect gases in the interval of temperature variation from the
critical temperature to the Boyle temperature is in good
agreement with the experiment if for the interaction between the particles
we consider the Lennard--Jones potential of the form
\begin{equation}\label{86_4_1}
U(r)=4\varepsilon\big(\frac{a^{12}}{r^{12}}-\frac{a^6}{r^6}\big),
\end{equation}
where $\varepsilon$  is the energy of the well depth and~$a$
is the effective radius. This agreement cannot be random.

The relations related to the microcanonical distribution
do not take pairwise interaction into account.
However, in the new axiomatics of the new ideal thermodynamics,
we take into account the form of this interaction,
namely, the Lennard--Jones potential, and the properties of this potential.

In the scattering problem with the Lennard--Jones potential,
the pair energy is of the form
\begin{equation}
E=4\varepsilon\big(\frac{a^{6}}{r^{6}}-\frac{a^{12}}{r^{12}}\big)\left(\frac{\rho^2}{r^2}
-1\right)^{-1}, \label{eq1a}
\end{equation}
where $\varepsilon$ is the depth of the well,
$a$ is the effective radius, and $\rho$ is the impact parameter.
Hence $r_0=\Phi(E,\rho)$. By replacing
$$
\frac ra=r' \qquad  \frac{\rho}{a}=\widetilde{\rho}
$$
we get rid of $a$.

For a given $\rho$, the minimum of $r_1$ and the maximum of $r_2$
on the graph in Fig.~\ref{fig1} are determined by the condition
\begin{equation}\label{aa}
\frac{dE}{dr}=0.
\end{equation}
At some point $\rho=\rho^0$, they coincide and, therefore,
\begin{equation}\label{bb}
\frac{d^2E}{dr^2}=0
\end{equation}
at some point $r^0$. On the graph of $E(r)$ (Fig.~\ref{fig1})
for $\rho=\rho_0$, the inflection points corresponds to $E=E_0$.
In general, we associate the point~$\rho$ with the value $E_{\max}$,
which is the maximal tangent to the well
and with the minimal values $E_{\min}$.
A pair of particles is ``incident from infinity'',
i.e., from the point $r=\rho$,
is tangent to the value $E_{\min}$,
and slips into the well of depth $E_{\max}-E_{\min}$.
Thus the actual well of scattering is turned over
with respect to the axis of the energy increase.
This is related to the fact that the attracting part
in the Lennard--Jones potential is negative.
The calculation yields $E^0 = 0.8\varepsilon$.
For $E>E^0$ the well (the trap) vanishes.

This criterion refers only to two colliding particles.
The usual argument in molecular physics involves the symmetry
of the average motion of the molecules in all six directions.
Therefore, $1/12$ of all particles move toward one another.
Since there are three axes, it follows that $1/4$ of all molecules
collide.\footnote{The isotropy principle
(of symmetry in all directions) is one of the key principles
of molecular physics. It must also be formulated rigorously
in mathematical terms as the isotropy principle
in the theory of Kolmogorov turbulence
or the Born--Karman conditions in the theory of crystals
(the problem of the crystal volume finiteness problem),
especially because of rapid development of the computer molecular dynamics
similar to the computer anisotropic turbulence~\cite{Kaneda1},~\cite{Kaneda2}.}.
Their mean energy is
$$
kT_{\text{B}} = \frac {16}{5}\varepsilon,
$$
where $k$ is the Boltzmann constant.
From physical considerations, the corresponding temperature, apparently,
corresponds to the so-called Boyle temperature\footnote{As
we shall see below, the ``well'' is overfilled at a lower temperature.}.

In Fig.~\ref{fig1}, which describes the behavior of the mass center,
the penetration beyond the barrier of particles moving from infinity and
scattering at one another is only possible in the quantum case.
However, in the presence of noise, such a penetration is also possible for
classical particles by the Pontryagin--Andronov--Vitt theorem
(see~\cite{Dykman}). In our case, by assumption, all the variants of the
microcanonical distribution are equiprobable. This condition yields significantly
larger amplitudes of deviation in white noise. Therefore, we can assume
that if the number of particles~$N$ exceeds some value of $N_{\text{cr}}$,
then particles penetrate the barrier under the condition that the
temperature is below~$T_{\text{B}}$ and transform into oscillating or
rotating pairs.

Setting $\frac ra=x$, \, $\frac{\rho}{a}=\widetilde{\rho}$, \,
we obtain
\begin{equation}
\label{g1}
E(x)=4\varepsilon\left(\frac{1}{x^{6}}-\frac{1}{x^{12}}\right)\left(\frac{\widetilde{\rho}^2}{x^2}
-1\right)^{-1}.
\end{equation}
The derivative of the function $E(x)$ is
$$
E'(x)=8\varepsilon
\frac{3x^8-2\rho^2x^6-6x^2+5\widetilde{\rho}^2}{x^{11}(-\rho+x)^2(x+\rho)^2}.
$$

After replacing $y=x^2$, the equation for the points of possible extremum
is of the form
$$
3y^4-2\rho^2y^3-6y+5\widetilde{\rho}^2=0.
$$
Hence
$$
\widetilde{\rho}^2=3y\frac{y^3-2}{2y^3-5},
$$
and the dependence
$$
E(y)=4\varepsilon\left(\frac{1}{y^{3}}-\frac{1}{y^{6}}\right)\left(\frac{\rho^2}{y}
-1\right)^{-1}
$$
is of the form
\begin{equation}\label{g2}
E(y)=4\varepsilon\left(\frac{1}{y^{3}}-\frac{1}{y^{6}}\right)\left(3\frac{y^3-2}{2y^3-5}
-1\right)^{-1}.
\end{equation}
The maximum value of $E_{max}=0.8$ is attained at the point
$y=\sqrt[3]{5}$.

The corresponding values of $\widetilde{\rho}$ are
\begin{equation}\label{g3}
\widetilde{\rho} =\frac{3}{\sqrt[3]{5}}.
\end{equation}

\begin{figure}
\begin{center}
\includegraphics{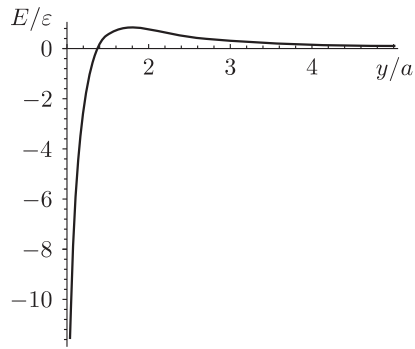}
\caption{The dependence of the energy $E$ on $y$} \label{fig2}
\end{center}
\end{figure}

The graph of the function $E(y)=E$, where $E<0$, has two points on the graph in
Fig.~\ref{fig2} for $y=y_1(E)<\sqrt[3]{5}$ and for $y=y_2(E)>\sqrt[3]{5}$.
These points correspond to the minimum and maximum
of the trap in Fig.~\ref{fig1}.

Note that the passage of a classical particle through the point~$y_2$,
when the energy of the particle corresponds to the point~$y_2$
(see Fig.~\ref{fig2}), is not a passage through an ordinary barrier.
In passing from quantum mechanics to classical mechanics, we see that
this point is no longer the usual turning point,
and the classical particle, in the limit, penetrates into the well
in infinitely large time.

In~\cite{TMF_159_1}, we already discussed Koroviev's trick
in a variety show when he scattered $n$ money bills among $k$
spectators and we shoed that if the number of spectators is twice
as large as some critical number~$k_0$, then, with a large probability,
one-half of the spectators does not get a single bill.
However, if they combine into pairs with a view to share the spoils, then
most of them will get some bills with a large probability.

From the point of view of Kolmogorov complexity, both scenarios are
equiprobable: none of them is preferable. As proved,
given such a canonical distribution, there is also a critical number
$N_{\text{cr}}$, as in the case of the Bose condensate.

By what has been said above, the axioms concerning the Bose distribution
that $N-N_{\text{cr}}$ particles pass to the zero level (stop)
are replaced by the following axiom: $N-N_{\text{cr}}$ particles form
dimers, trimers, and other clusters that do not affect pressure and form
Brownian particles in the gas\footnote{Here the author tried
to give a physical interpretation.
In mathematical theorems of the author, the behavior of dimers and clusters
is not considered as the behavior of physical objects.
Only the fractal dimension is determined, as well as
the partition of clusters into two classes.}.

The point~$E_{\text{cr}}$ for which the depth of the well (trap)
is maximum is $E_{\text{cr}}=0.286\varepsilon$. In our theory,
this point corresponds to the critical temperature
$T_{\text{cr}}=4\frac{E_{\text{cr}}}{k}$,
where  $k$ is the Boltzmann constant.

Let us present the table for
$0.286\frac{\varepsilon}{k}=\frac{T_{\text{cr}}}{4}$
for different gases for which the formula for the gas is
in correspondence with the depth $4\varepsilon$
of the well of the Lennard--Jones potential
(not to be confused with the depth of a trap in the
scattering problem).

\bigskip

\begin{center}
{\small
\begin{tabular}{|c|c|c|c|}
\hline
{Substance} & {$\varepsilon$, K} & {$T_{cr}/4$}  & {$E_{cr}\cdot\varepsilon/k$}\\
\hline
{$Ne$} & 36.3 & 11 & {10.5} \\
\hline
{$Ar$} & 119.3 & 37 & {35}\\
\hline
{$Kr$}& 171 & 52 & {50}\\
\hline
{$N_2$} & 95,9  & 31 & {28}\\
\hline
{$CH_4$} & 148.2 & 47 & {43}\\
\hline
{$C_2H_6$} & 243.0&  76 & {70} \\
\hline
{$C_3H_8$} & 242.0&  92 & {70} \\
\hline
{$C_4H_{10}$} & 313.0&  106 & {98} \\
\hline
{$CO_2$} & 213 &  76 & {52} \\
\hline
{$AsH_3$} & 281  &  93 & {82} \\
\hline
{$GeH_4$} & 237 &  77 & {59} \\
\hline
{$H_2S$} & 301 &  93 & {87} \\
\hline
{$H_2Se$} & 320 &  102 & {93} \\
\hline
{$NH_3$} &  300 &  101 & {87} \\
\hline
{$PH_3$} & 251.5  & 81  & {73} \\
\hline
{$SiH_4$} & 207.6 &  67  & {50} \\
\hline
\end{tabular}
}
\end{center}

\bigskip

By our ideology, the dependence $\frac{E_{\min}}{E_{\max}}$
on $E_{\max}$, corresponds to an isochore under constant volume.
The total energy equal to $E_{\max}$ corresponds to the temperature
and~$E_{\min}$ corresponds to the energy of the monomers.
In view of the fact that, by our scheme, only the monomers exert
chaotic pressure on the walls of the vessel,
$$
4E_{\min}=\frac{PV}{R}
$$
(the volume is measured in $\text{cm}^3/\text{mole}$).

Therefore,
$$
\frac{4E_{\min}}{4E_{\max}}=\frac{PV}{RT}=Z,
$$
where $Z$ is the compressibility factor.
The isochore $Z(P)$ is, indeed, the function
$\frac{E_{\min}}{E_{\max}}(E_{\max})$.

We study the temperature range from $T_{\text{cr}}$ to
$T_{\text{Boyle}}$, i.e., the domain called {\it fluids}
when the pressure exceeds $P_{\text{cr}}$ and the surface tension
making a liquid distinct from a gas, vanishes.

From the given table, we see that, for noble gases, the agreement
is to within several degrees, while, for more complex molecules,
it is to within 20 degrees on the Kelvin scale.
Nevertheless, the percentage is not so large, because the difference
of the temperatures, say, for water $H_2O$ and nitrogen $N_2$
is as much as several hundred degrees.

\begin{figure}
\begin{center}
\includegraphics{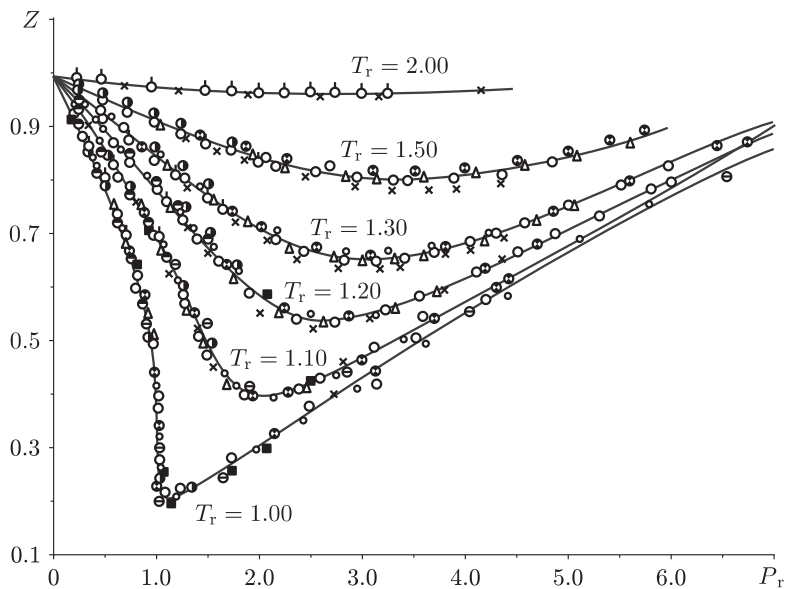}
\caption{Experimental data for different values of
$T_r=T/T_{\text{cr}}$ for different gases: $(\times)$~--- methane,
$(\circ)$~--- ethylene, $(\vartriangle)$~--- ethane,
$(\circlearrowright)$~--- nitrogen, $(\blacksquare)$~--- isopental,
$(\circleddash)$~--- $n$-heptane.
The other notation conventions:
small circle~--- $n$-butane,
vertically dashed circle~--- carbon dioxide,
circle dashed on the right~--- water.
The solid line corresponds to the data for hydrocarbonate.} \label{fig3}
\end{center}
\end{figure}

Let us present a graph in Fig.~\ref{fig3} for several similar gases
according to the law of corresponding states. We see that there is
a rather good agreement only for the isotherms corresponding
to the critical temperature, i.e., for
$T_{\text{r}}=\frac{T}{T_{\text{cr}}}$, where $T_{\text{r}}$
is the reduced temperature. The other isotherms give a mean
discrepancy of~5\%. Therefore, we can hope to obtain a better
agreement for the first isotherm at $T=T_{\text{cr}}$.

The principal thing for us is that the ideology involving
parastatistical condensate and the formation of dimers, not of points
located at the lowest level (as postulated in the case of the Bose gas)
has been fully corroborated by experiments.

On the graph in Fig.~\ref{fig3}, the critical point~$E=0.286\varepsilon$
corresponds to $Z=\frac{E_{\min}}{E_{\max}}=0.444$.
It is at this point that the avalanche process of parastatistical
condensate originate and it is further strenthened, because
the formation of dimers absorbs part of the energy, enlarges the condensate
and reaches the point at which all the monomers corresponding to $E_{\min}$
pass into rotating dimers.

This means that because the numbers of degrees of freedom of monomers
is equal to~$3$ and of rotating dimes is equal to~$2$, it follows that
this transition will occur up to the point $Z=\frac 23 0.444$,
equal to~$0.296$. In Wikipedia~\cite{Wiki}, the mean value for argon,
krypton, nitrogen, carbon dioxide, and methane equal to
$0.292$ is given. As we see, this value is in good agreement
with the graph in Fig.~\ref{fig3} and our concept of condensate
and dimers.

But, according to our scheme, dimers not only rotate. There is also a small
oscillatory component (the liquid is almost incompressible).
As pressure increases, this component also vanishes and, further, the
fluids behave as incompressible liquids and $Z(P)$ is a linear function.

The scattering problem for the  Lennard--Jones potential
divides the problem of the transformation of monomers into dimers
into two areas. The first area belongs to the case in which the height
of the barrier between the dimers and the monomers increases
and is considered up to its maximal height.
We say that this area of the scattering problem can be compared to the problem
of occurrence of dimers in abstract thermodynamics\cite{MZ_86_1}.

As the barrier decreases with decreasing temperature,
an additional energy barrier is needed to preserve the equilibrium
``monomers--dimers''. This follows from energy considerations.
When we speak of ``monomers'' and ``dimers'', we have in mind
a more general notion, because trimers and other clusters
may form. In addition, dimers can be in translational motion,
and thus contribute to the pressure on the walls.

We maintain that the equilibrium law well known in molecular physics for
sufficiently low pressures and relate to the number of atoms in the molecule
is violated in the sense of a greater role player by rotary
and oscillatory components owing to the occurrence of molecule association.
At the session of the Amsterdam Academy of Sciences in 1906,
Van der Waals said that the violation of the equation of state
discovered by him is due to the association of molecules.
In modern language, the term ``association of molecules'' means the formation
of dimers, trimers, and other clusters.

We divide the clusters into two forms:
1) those that contain at least one inner molecule (three-dimensional clusters)
and 2) those that have no such molecule.
The forme will be called ``domains'' by analogy with
the corresponding mathematical term.
Thus, domain are nanodrops, because they have molecules forming
the surface of the domain.
Among the gases noted in Fig.~\ref{fig2}
in the first part of the papaer~\cite{MZ_86_4}, there is carbon dioxide.
In it, micelles, repulsive monomers, are formed as well.
Thus, both dimers and micelles possess an additional barrier in the form
of surface tension.

The part of the scattering problem corresponding to the decrease
of the barrier with temperature in line with energy
considerations involves the appearance of domains and micelles
for the preservation of clusters and, therefore, another structure
(not yet studied by us) must appear in abstract thermodynamics.
We shall consider this situation in another paper.
However, we can hardly expect that the study of that situation will lead to
new discoveries in thermodynamics, possible only to the refinement
of the equations of state due to Van der Waals and other scientists.

In the present paper, we speak about a new phase transition
on the critical isotherm $T=1$ and at the critical pressure $P=1$.
The volume $V_{\text{cr}}/R$ undergoes a phase transition from
the value $V_{\text{cr}}/R=0.444$ to the value $V_{\text{cr}}/R=0.296$.
Here we deal with a phase transition of new type not discovered by
physicists despite the fact that it can be observed on experimental graphs
as a vertical segment (see Fig.~\ref{fig3} and~\ref{fig5}).
Nevertheless, the specialists in this area only say that this vertical line
is a result of the inaccuracy of the experiment forgetting about the fact that
in the ordinary Maxwell--Van der Waals phase transitions of the first kind,
the exact experimental data do not fully agree with the
theoretical calculations and the horizontal Maxwell segment cannot exist,
because the Avogadro number is finite.
This segment appears in theoretical constructions only in the
asymptotic limit.

The phase transition $ 0.444 \rightarrow 0.296$ corresponds to the
\textit{law of redistribution of energies}; actually, this effect
was discovered by the author in economics\footnote{As I mentioned
earlier, initially, I discovered a phase transition of this type
while studying economics crises, and later tried to discover
a similar effect in physics.}.

\begin{figure}
\begin{center}
\includegraphics{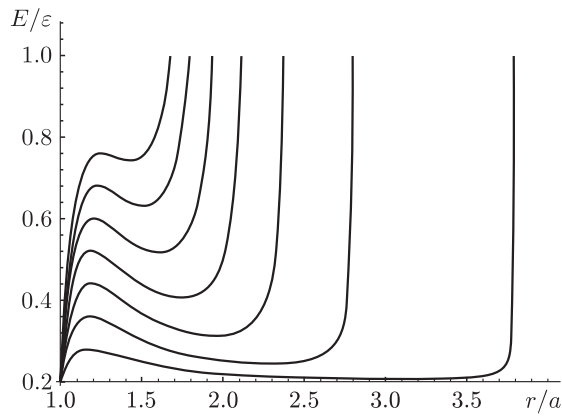}
\caption{Wells and barriers in the scattering problem of two particles
with Lennard--Jones interaction potential.} \label{fig4}
\end{center}
\end{figure}

The role of the points $E_{\min}$ and $E_{\max}$
in the scattering problem (see Fig.~\ref{fig4})
can easily be understood in terms of the general concept of the author
regarding microcanonical distributions constructed by the author.
The scattering problem is considered in the phase space:
$(r,\varphi,\theta)$ (the radius and two angles)
are the coordinates and $(p_r,p_{\varphi},p_{\theta})$
are the momenta. It can be reduced to a two-dimensional problem in which
$(p_{\varphi})$ is a free momentum. In addition, it has two invariants:
energy and moment, and hence can be reduced to a one-dimensional problem.
It is the points $E_{\max}$~\cite{ContMath} and the moment
equal to $E_{\max}\rho^2(E_{\max})$, where $\rho$ is the impact parameter
corresponding to the maximum of the well.

The relationship with the general concept will be elucidated in detail
in another paper. Here we present only arguments of physical nature.

The dimer can be formed in the classical domain if the scattering pair
of particles has energy equal to the height of the barrier.
This pair of particles penetrates into the well in ``infinite''
time and, due to the viscosity phenomenon and hence to a small loss
of energy, gets stuck in it for a while;
on the way back, this pair will hit the barrier as a result of the energy loss.
But if the pari of particles passes above of the maximum of the barrier, then it is
not known if there will be enough viscosity for this pair of particles to
hit the barrier. Therefore, it is the sliding point
that makes the main contribution\footnote{Since this pattern
occurs in the negative energy half-plane, it is more correct
to say `below the maximum'' instead of ``above the maximum''.
The meaning is the same, but the picture is more illustrative.}.

Because of viscosity, the fictitious particle reaches the well bottom
in a very large time, which means that only the rotational motion of the dimer
remains.
We have already mentioned the relaxation time influence,
explaining the deviation from the theoretical values
in the table of critical temperatures.
Just in the same way, the relaxation time affects
the oscillatory component. Therefore, it is natural
that the greater the deviation of actual critical temperatures
from their theoretical values,
the lower the ground value of the phase transition,
i.e., the larger the deviations from $Z=0.296$ of the values
given for the gases in the table.

\bigskip
\begin{center}
{\tiny
\begin{tabular}{|c|c|c|c|c|c|c|c|c|c|c|c|c|c|}
\hline
{Gas}&$Ne$&$Ar$&$Kr$&$N_2$&$CH_4$&$C_2H_6$&$C_3H_8$&$C_4H_{10}$&$CO_2$&$H_2S$&$H_2Se$&$NH_3$&$PH_3$\\
\hline
{$\frac{PV}{RT}$}&0.307&0.292&0.291&0.292&0.290&0.288&0.278&0.273&0.274&0.283&0.293&0.242&0.275 \\
\hline
\end{tabular}
}
\end{center}
\bigskip

The deviations of the data for propane $C_3H_8$, ammonia $NH_3$,
and carbon dioxide $CO_2$ from the theoretical data are the largest in the table.
Apparently, the ammonia and carbon dioxide to not reach the well bottom
because of the polarity effect.

It follows from abstract thermodynamics that if the dimension of
dimers and monomers is the same, the total energy inside and outside
the well is also the same.
Therefore, when we subtract the energy of dimers in the well it follows
from the total energy of the monomers, as is seen from our general concept,
that, for monomers, the energy remains equal to $E_{\min}=1/4 PV$.
The physical thermodynamic relations imply that such a levelling-off
occurs in infinite time. It is especially easy to see in the case
of a plasma when it is easy to reproduce a pattern
with a well and a barrier.

\begin{figure}
\begin{center}
\includegraphics{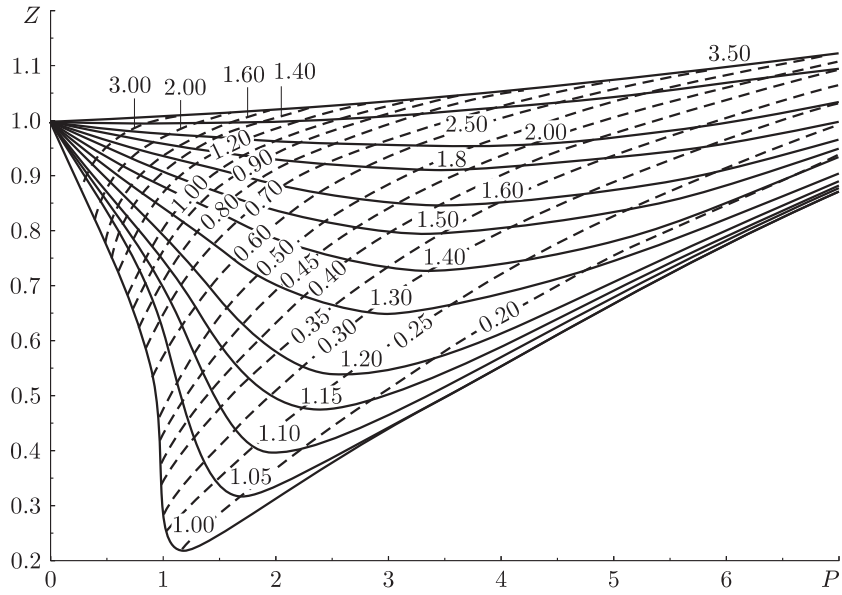}
\caption{Experimental graph.
$P=P_{\text{atm}}/P_{\text{cr}}$, $T_r=T/T_{\text{cr}}$,
$T$ is the temperature in Kelvin degrees,
$V/R$ is the volume in $\text{cm}^3/\text{mole}$,
$R$ is the gas constant, and  $Z= PV/(KT)$ is the compressibility factor.
The isochores $V/R= \text{const}$ are shown by dotted lines.}\label{fig5}
\end{center}
\end{figure}

\bigskip

\begin{figure}
\begin{center}
\includegraphics{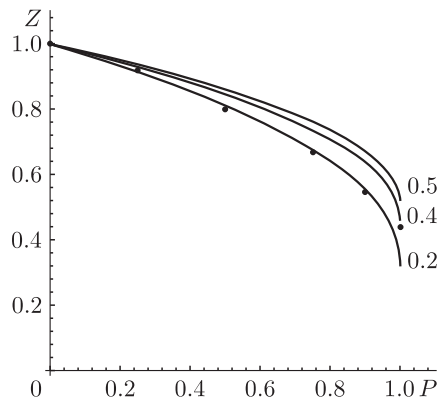}
\caption{Critical isotherm $Z(P)$ for dimensions~$3$ (curve~0.5),
$2.8$ (curve~0.4), and~$2.4$ (curve~0.2). Points correspond to
the experimental isotherm.} \label{fig6}
\end{center}
\end{figure}

Since the point $0.444$ corresponds to the fractal dimension~$2.8$,
and the point $0.296$ to the fractal dimension $2.45$,
using abstract thermodynamics~\cite{TMF_161_2},
we can compare the corresponding values
of $Z(P)$ for $P<1$ with experimental curves in Fig.~\ref{fig3}
and in Fig.~\ref{fig5}. As Fig.~\ref{fig3} shows,
experimental curves for different gases differ somewhat.
In Fig.~\ref{fig6}, we marked the values of the averaged isotherm
$T=1$ by black points.

In Fig.~\ref{fig5}, we demonstrate the law of corresponding
Van der Waals states and use the following notation:
$T_r= \frac{T}{T_{\text{cr}}}$,  $P_r= \frac{P}{P_{\text{cr}}}$.
Then the dimensionless quantity, i.e., the compressibility factor
$Z=\frac{PV}{RT}$ becomes a dimensional quantity in the units
$\text{cm}^3/\text{mole}$.
To avoid this, we assume that the volume~$V$
is also dimensionless, taking the ratio
$\frac{V}{R}\,\text{cm}^3/\text{mole}$ divided by unit volume
equal to $1\,\text{cm}^3/\text{mole}$ in Fig.~\ref{fig5}.
This volume is shown in the figure by dotted line marked by $1.00$.
Then all the quantities are dimensionless,
including the chemical potential.
We assume that $\mu_{\text{cr}}=T_{\text{cr}}$ and
$\mu_r=\frac{\mu}{\mu_{\text{cr}}}$.
In what follows, the subscript~$r$ is omitted.

First of all, we must find $Z(P)$.

To find the isotherm $T=T_{\text{cr}}=1$, we apply the well-known
(in thermodynamics) general formula for $T=\text{const}$,
$d\mu=V dP$, $V$ is the volume, $P$ is the pressure, and~$\mu$
it the dimensionless chemical potential.

Denote
$$
y=e^{\mu}, \quad 0 \geq\mu >-\infty
$$
as $\mu\to-\infty$, $y\to 0$.

Then, for an ideal gas (see~\cite{RJ_16_1})
\begin{equation}\label{eq1}
Z(y)=\frac{\Gamma(\gamma)}{\Gamma(\gamma+1)}\times
\frac{\int_0^\infty \frac{\varepsilon^{\gamma+1}\,
d\varepsilon}{e^{\varepsilon}-y}} {\int_0^\infty
\frac{\varepsilon^{\gamma}\,d\varepsilon}{e^\varepsilon-y}}.
\end{equation}
Therefore,
$$
d\mu=d\ln\,y=Z(y)d\ln\,P,
$$
and hence,
$$
\frac{d\ln\,y}{Z(y)}=d\ln\,P, \quad y=y(P).
$$

Denote $\ln\ P=\xi$, $\ln\ y=\mu$, $T\equiv 1$. Then
\begin{equation}\label{eq2}
\frac{d\mu}{Z(\mu)}=d\xi, \qquad
\xi=\int_0^\mu\frac{d\mu}{Z(\mu)}, \qquad
P=e^{\int_0^\mu\frac{d\mu}{Z(\mu)}}=e^{\xi}
\end{equation}
and $\mu=\mu(Z)$ according to formula~\eqref{eq1}.

As is known, in the $k$th-order parastatistics
of fractal dimension $\gamma$, the distribution over energy has the form
\begin{equation}\label{eq3}
Z_\gamma = \frac{\frac{1}{2m}\int_0^\infty
p^{2+\gamma}\bigg\{\frac{1}{e^{[(p^2/2m)-\mu]/\theta}-1}-\frac{k}{e^{[(p^2/2m)-\mu]k/\theta}-1}\bigg\}\,
dp} {\int_0^\infty
p^{\gamma}\bigg\{\frac{1}{e^{[(p^2/2m)-\mu]/\theta}-1}-\frac{k}{e^{[(p^2/2m)-\mu]k/\theta}-1}\bigg\}\,
dp}.
\end{equation}
For the Bose-statistics, $k=\infty$,
for the Fermi-statistics, $k=1$.
The parastatistics is characterized by a finite number~$k$.
In this case, the number of particles is assumed to be large.
We associated this number~$k$ with the number of particles
and obtain a new distribution:
$k=N$ for $T<1$ and $k=NT$ for $T>1$.
In our case, $k=N$,
where $N$ is the number of particles.
Although $N\to\infty$, the chemical potential $\mu$ still
can be positive and depend on~$N$.

We consider the parastatic energy distribution for $k=N$
in the case where $\mu\geq 1$ and the dimension is equal to
$2+2\gamma$ for $T=1$
\begin{equation}\label{eq3a}
\int_0^\infty \varepsilon^{1+\gamma}\big\{
\frac{1}{e^{-\mu}e^{\varepsilon}-1}
-\frac{N}{e^{-\mu N}e^{\varepsilon N}-1}\big\}\,d\varepsilon.
\end{equation}
Near the point $\varepsilon=\mu$ we expand the expression in braces
in a series up to $O\big((\varepsilon-\mu)N\big)^2$.
We have
\begin{align*}
\frac{1}{(\varepsilon-\mu)\big\{1+\frac{\varepsilon-\mu}{2}\big\}}
-
\frac{1}{(\varepsilon-\mu)\big\{1+\frac{N(\varepsilon-\mu)}{2}\big\}}
&=\frac{1}{\varepsilon-\mu}-\frac{1}{2+\varepsilon-\mu}
-\frac{1}{\varepsilon-\mu}+\frac{N}{2+(\varepsilon-\mu)N}
\\
&=\frac{N}{2+(\varepsilon-\mu)N}-\frac{1}{2+\varepsilon-\mu}.
\end{align*}
For $\varepsilon=\mu$, this expression tends to infinity as $N/2$.

Both terms in braces in~\eqref{eq3a} tend to zero. Obviously,
the integral
\begin{equation}\label{eq3b}
\int^{a?n}_{-a/N} \frac{N}{1+xN}\,dx=\int^{a/N}_{-a/N}
\frac{N\,dx}{1+\xi}=\ln(1+\xi)\big|^a_{-a}=\ln(1+a)-\ln(1-a)
=\ln\big(\frac{1+a}{1-a}\big)
\end{equation}
is finite for $a<1$.

At zero, the integrand exponentially tends to zero if
$\gamma>-1$ is integrable as $\varepsilon\to\infty$.
This implies that as $N\to\infty$,
the expression in braces tends to
$\delta(\varepsilon-\mu)$ and the integral in~\eqref{eq3a} tends to
$\mu^{\gamma+1}$. This implies that
$$
Z=\mu.
$$
And since
$$
d\mu=VdP=\frac{Z}{P}\,dP=\frac{\mu}{P}\,dP,
$$
for $P>1.5 P_{\text{cr}}$, up to $1/\ln N$ (i.e., up to~4\%),
we obtain the diagonal line on the graph $Z,P$
for the isotherm $T_{\text{cr}}=1$.
This corresponds to the strict incompressibility of fluid
as the pressure increases, which, as we see,
coincides with experimental graph in Fig.~\ref{fig5}.
For $\mu= \frac{\alpha\ln N}{N}$  ($\alpha=\text{const}$),
the graph of the critical isotherm decreases and becomes the diagonal
only for $\mu=\text{const}$.

The press $\mathbb{P}$ is an intensive quantity dual
to the fractal dimension~$\gamma$ for $\delta$,
$\mu>\delta>0$, independent of~$N$:
$$
\mathbb{P}=\frac{\partial\Omega}{\partial\gamma}=-\frac{V}{\gamma+1}
\int_0^\infty
\frac{\varepsilon^{\gamma+1}}{e^{(\varepsilon-\mu)/T}-1}
\big(\ln\,\varepsilon-\frac{1}{\gamma+1}\big)\,d\varepsilon
=\frac{VT}{\gamma+1}
\mu^{\gamma+1}\big(\ln\,\mu-\frac{1}{\gamma+1}\big).
$$

We note that the fluid critical volume at the phase transition instant,
a jump in the fractal dimension, is equal to $V_{\text{cr}}=0.444$.
This quantity is clearly determined. It also weakly depends
on the repulsion degree of the Lennard--Jones potential.
Therefore just this quantity is the main characteristics
of the critical volume. As was already said, it is rather
difficult to observe this quantity experimentally.
It is much easier to observe the lower end of the vertical segment.

If the trajectory of rotation of a pair of molecules
about the central point is nearly circular, then the energy
redistribution from translational to rotational
is $2/3V_{\text{cr}}=0.296$. For noble gases, the trajectory
ellipticity is rather small, and hence the energy redistribution
is $0.292$--$0.290$. But for carbon dioxide, it is more
``elliptic'', and hence the fall may attend the value~$2.7$,
but the original critical volume remains~$0.444$.

The energy redistribution law results in the phase transition,
see the vertical segment in Fig.~\ref{fig5}. Further, the condensate
is also formed as the fluid becomes incompressible in the end,
the fractal dimension is then~$2.4$. Although distribution~\eqref{eq3}
is similar to the Bose distribution, the final condensate is
an incompressible fluid consisting of clusters of dimension
less than three, i.e., it does not contain domains or micelles.

At higher pressures, clusters, i.e., nuclei of crystals, appear.
This fact also agrees with the author's general concept~\cite{TMF_161_2}.

Note that, for gas dimers, the fractal dimension is equal to $2.8$.
As is readily seen from arguments in~\cite{Landau},
if for the three-dimensional case, the Poisson adiabatic curve
is $PV^{5/3}=\text{const}$, then, for the dimension $2.8$, the adiabatic curve will
be $PV^{12/7}=\text{const}$, and, for ``liquid'' fluids of dimension
$2.4$ the adiabatic curve is of the form $PV^{11/6}=\text{const}$.

The averaged graph in Fig.~\ref{fig3} is of rather small accuracy, but this graph
in dimension $2.8$ is sufficiently well approximated
with the same degree of accuracy.

The point is that the experiment even with argon is highly unstable
bear the critical isotherm. Roughly speaking, it is considered as follows.
A cylinder with a freely sliding piston on its upper cover is first fixed
so that the volume is equal to the value of $V_{\text{cr}}$ for argon
$V/R=0.3$~$\text{cm}^3/\text{mole}$. Here part of the argon is in liquid state
and part of it in vapor state (see \cite{Strauf}).
Further, the cylinder is heated up from $86\,K$ to
$T=T_{\text{cr}}\approx159\,K$ until the surface tension film
disappears and a fluid is formed. So we obtain the lower point
of the vertical segment in Fig.~\ref{fig5}.
Further, slowly releasing the piston and supplying heat so that the
temperature remains equal to $T_{\text{cr}}$ all the time,
we must come to the point $P=0.25P_{\text{cr}}$.

If, as the result of the experiment, the temperature drops below
$T_{\text{cr}}$, then we have jumped to dimension $2.4$,
as is seen from the graph in Fig.~\ref{fig6}.
Therefore, here it is hard to abide by equilibrium thermodynamics,
while theoretical calculations yield the dimension~$2.8$\footnote{The
author believes that if the temperature is raised slightly
above $T_{\text{cr}}$ at first, and then the volume is increased
until the temperature drops to $T_{\text{cr}}$, then the dimension $2.8$
in the experiment will be preserved.
until $Z=0.9$, and then becomes $3$ in the phase transition of
the second kind (see below).}.

Essentially, after the redistribution of energies form the value $0.444$
to the value $0.296$, we come to a ``liquid'' fluid for all gases
for which the experimental graph holds.

The incompressible part of the critical fluid is obtained
for any dimension. And this variation in the dimension
is a very important new phenomenon.
The dimension reconstruction is a separate question.
Apparently, this phenomenon explains the well-known
``jamming'' effect for glass (cf.~\cite{Parisi}).
Thus, although the volume does not vary with pressure,
but, in this case, the dimers--clusters are ``chewed''
with a gradual reconstruction of clusters
in the direction of a more strict (ideal) architecture.
For the case of zero dimension and  the architecture ``self-organization'',
see~\cite{RJ_15_1}.

\begin{figure}
\begin{center}
\includegraphics{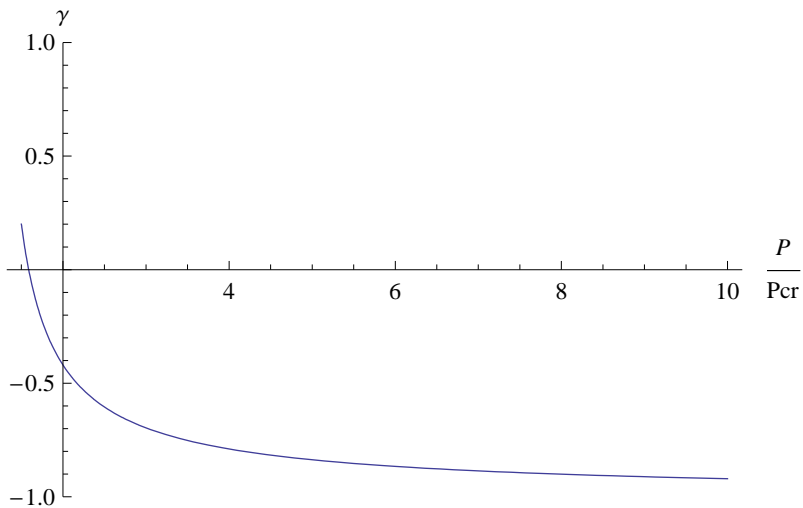}
\caption{The graph of decreasing fractal dimension $2(\gamma+1)$.} \label{fig7}
\end{center}
\end{figure}

The ``chewing'' effect, i.e.,
the dimension variation with~$P$ varying on the isotherm,
occurs according to the law
\begin{equation}\label{eq4}
\frac{d\gamma}{dP}=-\frac{\big[\frac{\partial
\Omega}{\partial\mu}\big]_{\mu=P}}{\frac{\partial\Omega}{\partial\gamma}},
\qquad
\Omega=-\int_0^{\infty}\varepsilon^{\gamma}\ln\big(1-e^{(\mu-\varepsilon)/T}\big)\,d\varepsilon,
\qquad \gamma|_{P=1.5P_{\text{cr}}}=0.2.
\end{equation}
The fractal dimension $\gamma$ decreases (is ``chewed'') by the law
(see Fig.~\ref{fig7})
$$
\frac{d\gamma}{dP}=(\gamma+1)\big(\ln\,P-\frac{1}{\gamma+1}\big)^{-1}.
$$

As we known, $\gamma=0.2$ corresponds to the fractal dimension~$2.4$.
Thus, the fractal dimension has the form $2+2\gamma$,
$\gamma>-2$. For $-1<\gamma\leq 0$, the parastatistical term~\eqref{eq3}, where $k=N$,
must be taken into account in~$\Omega$.

We described the critical isotherm $T=1$.
The process of passing to other isotherms is described
in~\cite{MZ_86_5},~\cite{TMF_161_3} by successive steps
of the $T$-mapping. In this paper, we do not consider this
procedure, which requires a lot of computer computations.

We consider the isochore
\begin{equation}\label{eq5}
\frac{E_{\min}}{E_{\max}}(E_{\min})
\end{equation}
for the scattering problem.
For a fixed~$\rho$, it means attraction, i.e., $|r_1-r_2|<\rho$,
where $r_1$ and $r_2$ are coordinates of two particles participating
in scattering and denoted by the letter $r=r_1-r_2$.
Thus, this is the problem of ``irradiation'' by a flow of monomers.
As was already mentioned, if $l_2=\rho-l_1<l_1$, where $l_1$ is the well width,
then the monomers get stuck in the well, and if $l_2=\rho-l_1>l_1$,
then a part of monomers return (i.e., are reflected).

In our case, the quantity $E^{(1)}_{\min}$
at this ``transition point'' is equal to~$0.536$,
and $E^{(1)}_{\max}=0.597$, and hence $Z\sim 0.9$.
In Fig.~\ref{fig5}, one can see that the isochore
$V/R=0.444\text{cm}^3/\text{mole}$ lying above this point
begins to stretch. This means that the isochore endures a
phase transition of the second kind.
This occurs due to variations in the dimension and, respectively,
in the entropy~\cite{TMF_161_2}. This problem, just as the relation
between the press and the pressure~$P$, must be studied separately,
because it is related to some generalization
of the basic thermodynamical notions~\cite{MZ_86_4}.

In the scattering problem, the volume is assumed to be equal to infinity.
According to nonstandard analysis, infinities can be graded.
In our case, we can consider the given isochore (Fig.~\eqref{eq5})
on different scales, simultaneously extending or shortening
both coordinate axes. It is only necessary to calculate the angle
which they make with the isotherm $T_r=1$.
Let us carry out these calculations.

For the isochore
$$
Z=\frac{E_{\min}}{E_{\max}}(E_{\min})
$$
at the point $V/R=0.444$, we can find the derivative at the point
$E_{\max}=E_{\max}^{\text{cr}}=0.286$.

We obtain
$$
\frac{dZ(E_{\min})}{dE_{\min}}\big|_{E_{\max}=E_{\max}^{\text{cr}}}
=a=1.951
$$

On the other hand, $Z=\frac{PV}{RT}$ and, for $T_r=1$,
$$
\frac{dZ}{dE_{\min}} =\frac{dZ}{d(PV)}.
$$
Hence
$$
\frac 1a= \frac{dPV}{dZ}=V\frac{dP}{dZ}+P\frac{dV}{dZ}
$$
\begin{align*}
\frac{dV}{dZ}\bigg|_{T=1}=\frac{1}{aP}-\frac{V}{P}\biggl(\frac{dZ(P)}{dP}\biggr)^{-1}&=
\frac{1}{aP}-\frac{Z}{P^2}\cdot\frac{1}{(dZ(P))/(dP)}= \\
\frac 1P\biggl(\frac 1a-\frac{Z}{P(dZ/dP)}\biggr)&=
\frac1P\biggl(\frac 1a-\biggl(\frac{d\ln Z(P)}{d \ln
P}\biggr)^{-1}\biggr).
\end{align*}

The value of $Z(P)$ was already obtained above \eqref{eq1}--\eqref{eq2}.

\begin{theorem}
The fractal dimension $\gamma$ does not vary along the isochore
up to a point at which the trap width in the scattering problem
for the interaction potential
coincides with the distance from the barrier maximum point
to the impact parameter~$\rho$
(Fig.~\ref{fig1}).
\end{theorem}

In our case, this coincidence occurs at the point
$Z=\frac{0.536}{0.597}\sim 0.9$. We note that,
for other Lennard--Jones potentials with a different degree of repulsion,
the point of phase transition of the critical volume
varies only in thousandth fractions.
The value $Z_{\text{cr}}$ at the critical point
$T_{\text{cr}}$, $P_{\text{cr}}$ is always equal to~$0.44$.

Thus, up to the value $Z=0.9$, we can reconstruct the family
of isochores, and hence of isotherms, because, at each point
of the isochore,~$Z$ and $P$, and hence~$T$, are determined.

Since in the three-dimensional case (the fractal dimension is~$3$)
$Z_{\text{cr}}$ for $\mu=0$ is attained at the point
$Z_{\text{cr}}=0.523$,
and above the point $Z=0.9$, the dimensions begin to increase
up to the three-dimensional, i.e., if there are no dimers
at the Boyle and higher temperatures,
then, according to the above argument,
the addition to the original number of particles--monomers is equal to
\begin{equation}\label{eq6}
\alpha\frac{l_2-l_1}{l_2+l_1}
\end{equation}
with a certain constant, which we determine by the condition:
for $\mu=0$, the dimension must be equal to~$3$,
i.e., $\gamma=1/2$.
Hence $\alpha=\frac{0.523}{0.444}-1=0.19$
for the Lennard-Jones type potential~\eqref{86_4_1}.

We find $\gamma$ from the condition
$$
Z_{\gamma}\big|_{T=1,\mu=0}=0.444+0.19\frac{l_2-l_1}{l_2+l_1}\cdot 0.444.
$$
Now, in our case, $\gamma$ varies from $0.4$ to $0.5$
and depends on~$E_{\min}$, i.e, on $\frac{PV}{4}$, where $V=0.444$.
But
$$
-PV=T^{2+\gamma(PV)}\int_0^\infty \frac{\varepsilon^{1+\gamma}
d\varepsilon}{e^{\varepsilon-\mu(T)}-1},
$$
where $\mu(T)$ is determined by the isochore
$\frac{E_{\max}}{E_{\min}}(E_{\min})$ continued above
the point $Z=0.9$.
Now we have determined the dependence $Z(P)$ for the new isochore
whose dimension varies from~$2.8$ to~$3$ and the volume remains
unchanged.

The quantity $l_2-l_1$, as a function of temperature,
increases almost exponentially approaching the Boyle temperature.
To retain this fast increasing number of monomers in the volume~$0.444$,
it is necessary to increase the pressure at the same rate.
Therefore, the pressure increases much faster than~$Z$,
and this results in elongation of the isochore for $Z>0.9$.
The value of~$Z$ increases by~$0.19$ units and attains $Z=1.19$,
while the pressure increases more than twice.

The other isochores are similar to those constructed above,
because their abscissas and ordinates increase and decreases
by the same factors.
The angle at which they intersect the critical isotherm
$T=1$ was calculated. To each point in the family of isochores
there corresponds a pair of points $Z=\frac{PV}{RT}$ and~$P$.
Hence, for a given $V$, the temperature~$T$ is determined
and the locus of points corresponding to the isotherm $T=\text{const}$
is constructed.

In conclusion, I note that the rigorous proof
of all the above statements is very cumbersome.
It is based on the use of the correction to variations
in the self-consistent field in the scattering problem
and the complex germ method
(in topology, the term ``Maslov gerbe'' is used,
in mathematical physics, the term ``complex germ''),
as well as the ultrasecond quantization,
where the operators of couple creation are used.

A more precise method of proof
(the method of $T$-mappings) is given in~\cite{TMF_161_2},~\cite{TMF_161_3}.

The author wishes to express his deep gratitude
to A.~R.~Khokhlov, A.~A.~Kulikovskii,
A.~I.~Osipov, I.~V.~Melikhov, L.~R.~Fokin, Yu.~A.~Ryzhov,
A.~E.~Fekhman, A.~V.~Uvarov, P.~N.~Nikolaev, M.~V.~Karasev, and
A.~M.~Chebotarev for fruitful discussions and also thanks
A.~V.~Churkin, D.~S.~Golokov, and D.~S.~Minenkov for help
in computer computations.

\end{document}